\documentclass[prl,aps,twocolumn,groupedaddress,showpacs,floatfix]{revtex4}
\usepackage{amsmath,amssymb,multirow,epsfig,bm}
\newcommand{\gsim}{\, \raisebox{-0.8ex}{$\stackrel{\textstyle >}{\sim}$ }}
\newcommand{\lsim}{\, \, \raisebox{-0.8ex}{$\stackrel{\textstyle <}{\sim}$ }}
\newcommand{\etal}{{\it et al.,\;}}
\newcommand{\beq}{\begin{equation}}
\newcommand{\eeq}{\end{equation}}
\newcommand{\bea}{\begin{eqnarray}}
\newcommand{\eea}{\end{eqnarray}}

\newcommand{\benn}{\begin{displaymath}}
\newcommand{\eenn}{\end{displaymath}}
\newcommand{\ket}[1]{| #1 \rangle}                     

\begin{document}

\title{Polarization Measurements and the Pairing Gap in the Universal Regime}

\author{ J. Carlson and Sanjay Reddy }
\affiliation{Theoretical Division, Los Alamos National Laboratory Los Alamos, NM 87545,  USA}

\begin{abstract}
We analyze recent cold-atom experiments on imbalanced Fermi systems using a minimal model in which a BCS-like superfluid phase coexists with a normal phase.  This model is used to extract the T=0 pairing gap in the fully paired superfluid state. The recently measured particle-density profiles are in good agreement with the theoretical predictions obtained for the universal parameters from previous Quantum Monte Carlo (QMC) calculations. We find that the zero-temperature pairing gap is greater than 0.4 times the Fermi energy $E_F$, with a preferred value of $0.45 \pm 0.05$ $E_F$. The ratio of the pairing gap to the Fermi Energy is larger here than in any other system of strongly-paired fermions in which individual pairs are unbound.   
\end{abstract}

\date{\today}


\maketitle
Experiments that trap and cool fermionic atoms are now providing new insights on the nature of strongly-interacting Fermi systems with a number imbalance between the interacting species \cite{Partridge:2006b,Shin:2006,Partridge:2006a,Zwierlein:2006,Shin:2007}.  Experiments to date have studied systems containing two hyperfine states of $^6$Li, which we label $\ket{\uparrow}$  and $\ket{\downarrow}$ for convenience.
These experiments can tune both the number asymmetry (polarization) and the interaction strength and thus have the potential to probe new phases of superfluid matter that are expected on theoretical grounds. Furthermore, they can directly probe pairing gaps in these strongly-paired superfluid systems. 
In this paper we examine recent experimental data on polarized Fermi systems and extract the T=0 pairing
gap in the BCS phase.  The ratio $\delta$ of the pairing gap to the Fermi energy at unitarity is a universal parameter and previous QMC calculations predict $\delta \approx 0.5$.  Gaps in strongly-paired Fermi systems are  important
in diverse areas of physics including nuclear matter in neutron stars and the phase structure of dense QCD.


Cold-atom experiments probe the strongly-interacting regime where the superfluid properties are robust at finite temperature. When the short-range interaction is tuned to produce an infinite scattering length all measurable quantities are related to their free Fermi gas counterparts by universal constants since the interaction does not present a dimensionful  scale.  For this reason,  the system with $a=\infty$ is said to be in the  ``universal" regime.  
In this regime two phases are certain to exist: a superfluid state at zero polarization and a normal state at
large polarization.  The superfluid state at T=0 can be characterized by the ground-state energy  $E_{SF} = \xi (3/5) E_F $ and the pairing gap $\Delta = \delta E_F$
with $ E_F = ( 3 \pi^2 \rho)^{2/3}/2m$.  The normal state can be characterized by the binding of a minority spin particle to
the Fermi sea of majority particles, $E_N - E_0 = - \chi E_{f,N}$, where $E_N$ is the energy of the normal state, $E_0$ is the energy of non-interacting
particles, and $E_{f,N}$ is the Fermi energy of the majority spin population $E_{f,N} = ( 6 \pi^2 \rho)^{2/3}/2m$. 

The universal constants have been calculated using Quantum Monte Carlo techniques,\cite{Carlson:2003,Giorgini:2004,Chang:2004,Carlson:2005,Lobo:2006} 
yielding: $\xi$ = 0.42(01), $\delta = 0.50(03)$,
and $\chi = 0.60(.01)$.  More recent calculations of $\xi$ using both Diffusion Monte Carlo and Auxiliary Field Monte Carlo \cite{Carlson:2007} indicate
that $\xi$ may be slightly smaller, $\xi = 0.40(.01)$.
While these calculations are approximate, calculations of the respective ground-state energies are upper bounds 
and hence provide an apparently accurate upper bound to the parameter $\xi$ and lower bound to $\chi$ \cite{Umass:2007}.  The pairing gap parameter
$\delta$ is obtained from the difference between even- and odd-particle number simulations, and hence is not a bound on the true value.
Measurements of the parameters $\xi$ and $\chi$ appear to be consistent within errors with these calculations as shown below.

These universal parameters are defined precisely at polarization zero and one for the superfluid and normal states, respectively.
QMC calculations of the superfluid and normal state suggest that simple descriptions based upon a quasiparticle
picture and these universal parameters are valid over a wide range of polarizations between zero and one.  At $T=0$ the superfluid + quasiparticle
picture appears to work reliably up to polarization $ \sigma  =(  n_{\uparrow} - n_{\downarrow} )/ (  n_{\uparrow} + n_{\downarrow} ) \leq 0.2$ \cite{Carlson:2005}, while in the normal state the simple quasiparticle picture appears to work reliably for polarizations $\sigma \geq 0.4$ \cite{Lobo:2006}. The phase transition between these two states is first-order at low temperature.  At T=0 the transition occurs between an unpolarized superfluid state and a normal state at a polarization of approximately 0.4 \cite{Lobo:2006}.  At the transition $\delta\mu = \delta\mu_c \simeq \mu$ where $\delta\mu=(\mu_\uparrow - \mu_\downarrow)/2$ and $\mu=(\mu_\uparrow+\mu_\downarrow)/2$ \cite{Carlson:2005,Lobo:2006}.  Here the spin-up and spin-down particle chemical potentials are $\mu_\uparrow$ and  $\mu_\downarrow$, respectively. The pairing gap
is only slightly larger than the value required for the stability of a gapless superfluid with finite polarization at zero temperature.
At finite temperature, however, the  superfluid quasiparticles will be thermally excited, polarizing the superfluid; 
this polarization can be exploited to extract the pairing gap in the superfluid. The main goal of this study is to 
extract these fundamental properties of the system from experiments in the universal regime. 

 The first-order phase transition
appears to have been observed in recent MIT \cite{Shin:2006} and Rice \cite{Partridge:2006a} experiments.  Significant differences
remain, however. In particular the Rice experiment finds a transition directly from zero to full polarization.  Such a scenario cannot
happen in a bulk system if the binding of spin-down particles in the normal state ($\chi = 0.6$) is so large, and hence the 
observed polarization must be related to the unique geometry of the trap. The MIT experiment has been performed for larger 
numbers of particles and more spherical traps, and hence the simple local density approximation employed here is better suited for an analysis 
of these experiments. 

Our picture of the two states at very low temperatures are presented below.  The calculated universal parameters then yield precise
density distributions for the minority and majority species in the trap.


\underline{Superfluid state with a small polarization:} In the universal regime the energy density, pressure and the chemical potential of the unpolarized superfluid state are given by $\epsilon_{SF} = \xi~k_F^5/(10 \pi^2 m)$, $P_{SF} = \xi~k_F^5/(15 \pi^2 m)$ and $ \mu=\xi ~k_F^2/2m$, respectively. Here $k_F$ is the  Fermi momentum  $k_F=(3\pi^2 \rho)^{1/3}$.  In Ref.~\cite{Carlson:2005} we calculated the excited state quasi-particle (qp) dispersion relation by introducing additional fermions to the unpolarized superfluid state. To leading order in the momentum expansion, the dispersion curve for qp's measured relative to $\mu$ is given by 
\begin{equation} 
\omega_{\rm{qp}}(k)=\Delta \sqrt{(1+a_2~(x-x_0)^2 + {\cal O} [(x-x_0)^4)]} \,,
\end{equation}      
where $x=k^2/k_F^2$, $x_0 \simeq 0.83(3)$, and $a_2 \simeq 1.3(2)$. The higher order terms are negligible at small polarization.  

The energy to excite a quasiparticle in the vicinity of the normal-superfluid transition is anomalously small in strong coupling since $\delta\mu_c \sim \Delta$ at the first-order transition \cite{Carlson:2005}.  Thus even when $T \ll E_F$  spin-up quasiparticles are thermally excited in the superfluid in the vicinity of the transition. Further, zero-temperature QMC calculations suggests that 
interactions between quasi-particles in the superfluid are weak at low polarization. Hence the number density and pressure of thermally excited quasi-particles are taken to be
\begin{eqnarray} 
n_{\rm qp} &=& \int \frac{d^3k}{(2\pi)^3}~\left(1+\exp\left(\frac{\omega_{\rm qp}(k)-\delta\mu}{T}\right) \right)^{-1}\,,\\
P_{\rm qp} &=& T \int \frac{d^3k}{(2\pi)^3}~{\rm Log}\left[1+\exp\left(\frac{\omega_{\rm qp}(k)-\delta\mu}{T}\right) \right] \,,
\end{eqnarray} 
respectively. The superfluid phonons with a dispersion relation $\omega= c~k$, where $c= \sqrt{\xi/3}~ k_F/m$, are negligible compared to spin-up quasi-particles except at very low temperature when $T \ll (\Delta-\delta \mu)$.    
 
We anticipate that the gap will decrease due to both finite temperature and polarization.   For the present calculations we assume that for low temperature the pairing gap decreases slowly according to $\Delta(T) = \Delta(0) \sqrt{ 1 - (T/t_c~E_F)^2)}$. The zero-temperature pairing gap is a constant fraction $\delta$ of the Fermi energy: $\Delta (0) = \delta E_F$. The coefficient $t_c$  controls the temperature dependence of the gap and depends on $\Delta (0)$ and the ratio $\delta\mu/\mu$. Its calculation is beyond the scope of this work and a mean-field treatment of the polarized finite temperature superfluid phase is discussed in Ref.\cite{Stoof:2006}. Here we treat $t_c$ as a parameter in the range $0.05 - 0.25$.  

\underline{Normal state at high polarization:}
Through QMC studies we have determined the energetics of the normal state at high polarization. Near unit polarization, earlier work has shown that a mean field shift for the spin-down particle given by 
\begin{equation} 
E_{\downarrow}(k)=\frac{k^2}{2m} - \chi~\frac{k_{F_\uparrow}^2}{2 m } 
\label{ref:asym_nrml}
\end{equation} 
with $\chi=0.6$ provides a good description of QMC results for the energy at zero temperature \cite{Lobo:2006} . To extend the analysis to finite temperature we  adopt a simple independent-particle model in which the interactions modify the single-particle levels and are able to fit the  QMC results. A symmetric form for the dispersion relation that fits the QMC results is given by 
\begin{eqnarray}
E_{\uparrow}(k)=\frac{k^2}{2m} - \chi~\frac{k_{F_\downarrow}^3}{4 m \tilde{k}_F}; \quad
E_{\downarrow}(k)=\frac{k^2}{2m} - \chi~\frac{k_{F_\uparrow}^3}{4 m \tilde{k}_F} 
\label{ref:sym_nrml}
\end{eqnarray} 
where $\tilde{k}_F=(k_{F_\uparrow}^6+k_{F_\downarrow}^6)^{1/6}$ and $\chi=0.6$.   We note that the dispersion relation in Eq. \ref{ref:asym_nrml} and the symmetric form in Eq. \ref{ref:sym_nrml} both provide a fair description of the QMC data near unit polarization and differ only by a few percent at $\sigma=0.5$.

\begin{figure}
\vspace{-0.2in}
\begin{center}
\includegraphics[height=3.5in,angle=-90]{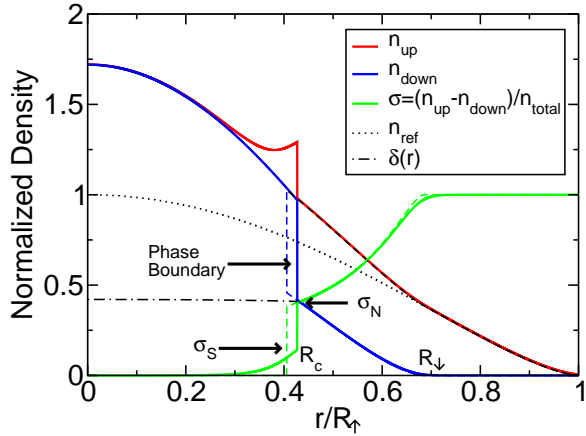}
\caption{Spin-up and spin-down densities and polarization versus radius predicted by theory for $\delta = 0.43$ at $T'=0$ (dashed)  and $T' = 0.03$ (solid) are shown. The  universal parameters $\xi$ and $\chi$ are taken 
as 0.4 and 0.6, respectively. For $T'=0.03$ the local value of $\delta$ shown by the dot-dashed curve was obtained using $t_c=0.1$. }
\label{fig:density-profile}
\end{center}
\vspace{-0.2in}
\end{figure}

We assume a harmonic trap and work in the local density or Thomas-Fermi approximation. The physical state at any location in the trap is completely determined by the local chemical potentials $\mu_{\uparrow}=\mu+\delta \mu$ and $\mu_{\downarrow}=\mu-\delta \mu$. The chemical potential $\mu = \lambda -V_{\rm{Trap}}(r)$, where $V_{\rm{Trap}} (r)= 0.5~\hbar \omega~(r/r_0)^2$ and $r_0=\hbar c / \sqrt{\hbar \omega ~m c^2}$ while $\delta \mu$ is constant since the trap does not distinguish between the different fermion species. Using the notation of reference \cite{Shin:2007} distances are scaled relative to the radius $R_{\uparrow}$  at which the zero-temperature spin-up density vanishes. The radius $R_{\downarrow}$ is similarly defined as the radius at which the zero-temperature spin-down density vanishes. The temperature $T'$ is scaled to the  Fermi energy of a non-interacting Fermi gas at $r=0$ with the same density profile as the exterior cloud. Similarly the densities are measured in units of $n_0$, where $n_0$ is the density of a non-interacting Fermi gas at $r=0$ with the same density profile as the exterior cloud. At low temperature the density profiles predicted by the theory discussed above are shown in Fig. \ref{fig:density-profile}.  The inner superfluid region and the outer normal region separated by a first-order transition are clearly seen. The transition is characterized by $\sigma_s$ and $\sigma_n$ -- the polarizations in the superfluid and normal phase, respectively.  As mentioned earlier, at low temperature $\sigma_N \simeq 0.4$ while $\sigma_s$ increases rapidly from zero at zero-temperature with increasing temperature due to the low excitation-energy of spin-up quasiparticles in the vicinity of the transition. In our implementation of the equation of state for the normal phase, thermal effects on the density profiles are negligible except close to $r=R_\downarrow$ and  $r=R_\uparrow$. 

The density distribution of the spin-up particles (majority species) between $R_{\downarrow}$ and $R_{\uparrow}$  provides a measure of the temperature and the chemical potential
of the spin-up particles. Similarly, the density distribution of the unpolarized superfluid state at the origin provides an independent measure of the 
average chemical potential $\mu$ assuming the calculated value of $\xi$. Since $\chi=0.6$ describes the normal state near unit polarization, the radius $R_{\downarrow}$ can independently provide a measure of the spin-down chemical potential. Thus, the two radii, the density at large radii,
and the central density measure the two chemical potentials and the temperature and provide a consistency check between the calculated values of $\xi$ and $\chi$ in extracting $\mu$ and $\delta \mu$.

In order to compare with experimental results,  we assume a specific ratio of the density at the origin to the maximum density of a cloud of free fermions with the same density distribution for $r > R_{\uparrow}$,
which is equivalent to assuming a specific average chemical potential $\mu$.
The values of the ratio $R_{\downarrow}/R_{\uparrow}$, the total polarization $P_{tot}$ in the trap, the transition radius $R_c$, and the polarization
as a function of radius are then predicted for various values of the pairing gap and the temperature.  We find that present measurements
provide strong constraints on the pairing gap even though the temperature is not very precisely determined in the lowest
temperature measurements.

\begin{figure}
\vspace{-0.2in}
\begin{center}
\includegraphics[height=3.5in,angle=-90]{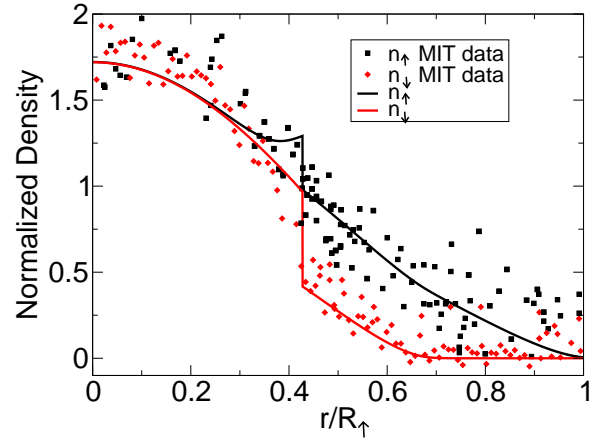}
\caption{Spin-up and spin-down densities theory and experiment for $\delta = 0.43$, $\xi=0.4$ and $\chi=0.6$ at $T' = 0.03$.}
\label{fig:compare-densities}
\end{center}
\vspace{-0.2in}
\end{figure}

Fig. \ref{fig:compare-densities} compares the calculated spin-up and spin-down densities as a
function of radius. We used a normalized temperature $T'=0.03$ and a normalized density at the origin $n_\uparrow(r=0)/n_0= n_\downarrow(r=0)/n_0 =1.72$ consistent with the  experiment. This simple model reproduces the radius of the transition, the radius where the spin-down density goes to zero, and
the overall polarization in the trap.   The calculated polarization is 0.44 for these values of the parameters,
and the measured value is 0.44(0.04) \cite{Shin:2007}.

\begin{figure}
\begin{center}
\vspace{-0.2in}
\includegraphics[height=3.5in,angle=-90]{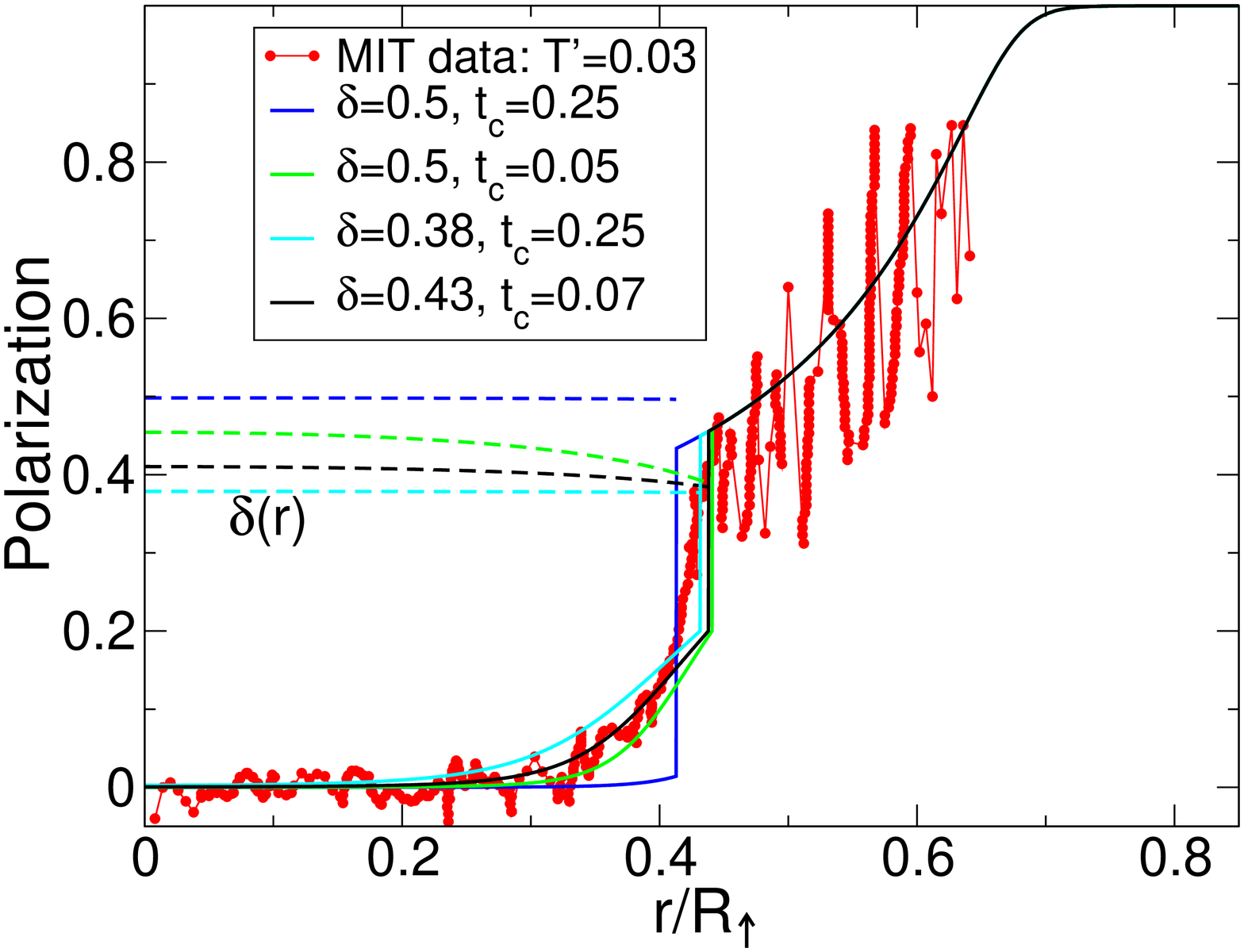}
\includegraphics[height=3.5in,angle=-90]{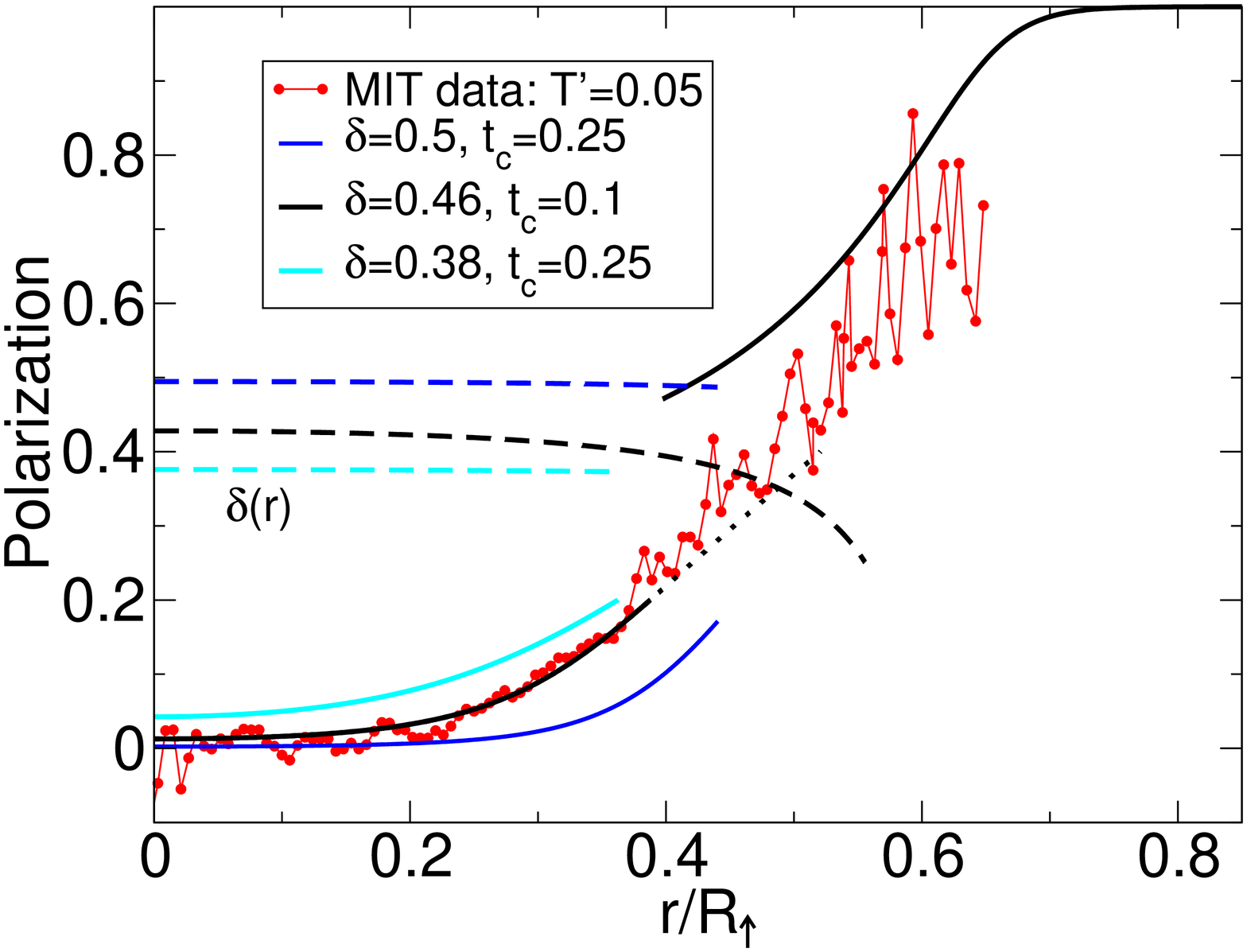}
\caption{Polarization versus radius, theory and experiment, for different values of $\delta$ and $t_c$ at $T'=0.03$ and $T'=0.05$. The dashed curves show the local finite temperature gap. The results indicate that the data provide both an upper and a lower bound on the gap: $0.5 \ge \delta \ge 0.4$. }
\label{fig:compare-polarization}
\end{center}
\end{figure}

In Fig. \ref{fig:compare-polarization} we compare the measured and calculated polarization at $T'=0.03$ and $T'=0.05$. Results for different values of $\delta$ and $t_c$ are shown. At $T'=0.03$ the polarization is approximately $0.12$ at the interface at $r/R_{\uparrow} = 0.43$, consistent with the experimental results. The qualitative and quantitative features of the measured polarization at $T'=0.03$ are captured by the normal phase  at $r/R_\uparrow \gsim 0.45$ and a thermally polarized superfluid phase for $r/R_\uparrow \lsim 0.4$ -- consistent with a first-order transition somewhere in between. In contrast, the comparison between theory and data at  $T'=0.05$ suggests that the superfluid extends further out.  Extrapolation of the polarization predicted by theory in the superfluid state (dotted curve) to $p\simeq 0.4$ provides a much better description of the data than the normal state prediction and  a clear signature of a first-order transition is absent. In both cases there appears to be evidence for a mild decrease in the gap with increasing $T/E_F$ and polarization. 

For a fixed central density and $R_\uparrow$, our minimal model predicts that the phase-boundary $R_c$ moves outward in the trap with increasing temperature. This behavior is very sensitive to the thermal properties of both phases at low temperature. In our analysis, which is expected to hold only for small temperature and polarization, the thermal response of the superfluid phase in the vicinity of the transition is stronger than that of the normal phase -- driven entirely by the fact that spin-up quasiparticles are easy to excite and have a large density of states.   

The comparison in Fig.~\ref{fig:compare-polarization} provides compelling lower and upper bounds
for the superfluid gap. Even if the temperature was extracted incorrectly from the experiment, the extracted gap cannot
be too small.  A gap smaller than $\approx 0.4 E_F$ would produce a shell of polarized superfluid before the
transition even at zero temperature.  Furthermore, the radial dependence of this polarization would be quite different
than observed experimentally, rising abruptly from the point where $\Delta = \delta \mu$ and 
being concave rather than convex. A gap larger than $\approx 0.5 E_F$ would be unable to produce the observed polarization in the superfluid phase.

We have also examined the dependence of our results on the universal parameters $\xi$ and $\chi$.  Both of these are expected to
be uncertain by $0.02$.  These uncertainties, as well as the uncertainties in the superfluid quasiparticle
dispersion relation do not significantly alter the extracted bounds on the superfluid gap.  If the transition temperature and gap decrease significantly at finite temperature with increasing polarization, the extracted gap could be slightly higher.


We note that the pairing gap extracted here is significantly larger than that obtained in simple analysis of RF spectroscopy.
These experiments measure the response to a probe tuned to a transition between a minority spin, for example, and a third
hyperfine state \cite{Gupta:2003,Chin:2004,Shin:RF2007}.  RF spectroscopy analysis, however, is complicated by the strong final-state interactions that reduce the average
energies of the transition\cite{Baym:2007}. 

In summary, it is possible to extract the pairing gap from measurements of polarized Fermi gases in the unitary regime.
These systems have an extremely large gap of almost one-half the Fermi energy -- the value extracted in this work is clearly the largest 
gap measured in any Fermi system. Precise measurements of the particle densities in the unitary regime, and measurements extending into the BCS regime would be valuable, in particular as direct experimental tests of the pairing gap in low-density neutron matter which is relevant in  neutron stars.

We would like to thank M. Alford, A. Gezerlis and Y. Shin for useful comments on the manuscript.
The work of S.R. and J.C. is supported by the Nuclear Physics Office of the U.S. Department of Energy and by the LDRD program
at Los Alamos National Laboratory.


\end{document}